# Experimental Study of Acceptor Removal in UFSD[1]

Y. Jin[a], H. Ren[a], S. Christie[a], Z. Galloway[a], C. Gee[a], C. Labitan[a], M. Lockerby[a],
F. Martinez-McKinney[a], S. M. Mazza[a], R. Padilla[a], H. F.-W. Sadrozinski[a*], B. Schumm[a],
A. Seiden[a], M. Wilder[a], W. Wyatt[a], Y. Zhao[a], R. Arcidiacono[b,c], N. Cartiglia[b], M. Ferrero[c],
M. Mandurrino[b], F. Siviero[d], V. Sola[b,d], M. Tornago[d] V. Cindro[e], A. Howard[e], G. Kramberger[e],
I. Mandić[e], M. Mikuž[e]

[a]*SCIPP, UC Santa Cruz, Santa Cruz, CA 95064, USA*
[b]*INFN, Torino, Italia*
[c]*Universita del Piemonte Orientale, Italia*
[d]*Universita del Studie Torino, Torino, Italia*
[e]*Jozef Stefan Institut (JSI), Dept. F9, Jamova 39, SI-1000 Ljubljana, Slovenia*

**Abstract**

The performance of the Ultra-Fast Silicon Detectors (UFSD) after irradiation with neutrons and protons is compromised by the removal of acceptors in the thin layer below the junction responsible for the gain. This effect is tested both with capacitance – voltage, C-V, measurements of the doping concentration and with measurements of charge collection, CC, using charged particles. We find a perfect linear correlation between the bias voltage to deplete the gain layer determined with C-V and the bias voltage to collect a defined charge, measured with charge collection. An example for the usefulness of this correlation is presented.

## 1. Introduction

A recent advance in semiconductor particle detectors are Ultra-Fast Silicon Detectors (UFSDs) which reach single particle timing resolution of 10-20 ps [1]. They consist of thin (20 – 50 µm) n-in-p Low-gain Avalanche Detectors (LGADs) which have internal gain of 10-20 due to a highly doped p++ layer between the high resistivity p-bulk and the junction [2, 3, 4]. The internal gain improves the signal-to-noise ratio S/N such that excellent time resolution in thin sensors is achievable [5]. The gain is governed by a high electric field which depends on both the doping profile of the gain layer (the quantity as well as the spatial distribution) and the applied bias voltage of the sensor.

In this paper, the correlation between parameters of the doping profile from Capacitance-Voltage (C-V) measurements and the bias dependence of the gain is explored. The doping profile is characterized by the voltage to deplete the gain layer, $V_{GL}$, and the gain by the bias V(G=x) at which a gain of G=x is reached .

This correlation can be employed for quality assurance of the data as well as for predicting the performance at higher bias which can not be reached with the present sensors. The method is explained first on an un-irradiated set of four LGADs produced by Hamamatsu Photonics (HPK) of 50 μm active thickness with doping concentration changing in steps of 10% each [6]. The highly doped p++ layer is created by boron implantation for all the sensors studied in this paper. Since these sensors all have the same geometry of the gain layer we can explore the effect of different amounts of doping as the only variable. Comparing a sensor with more initial doping after irradiation to an unirradiated sensor with lower initial doping also allows a direct determination of the amount of dopant reduction with fluence.

These measurements are then extended to irradiated sensors from HPK and FBK (Foundation Bruno Kessler) with distinctly different properties of the gain layer, including infusion of Carbon. The observed change of the doping density and the required increase in V(G=8) is then interpreted as caused by a process called "Acceptor Removal" where the dopant in the gain layer is increasingly de-activated by the rising concentration of interstitials on the Si lattice [7, 8]. Since the gain is lower after irradiation, the required gain is lowered from G=10 in the pre-rad case to G = 8 after irradiation.

The observed reduction in gain results from the reduced electrical field in the gain layer. The measurement of several sensor types allows us to evaluate the effect of the different doping profiles as well as Carbon infusion.

## 2. Doping and Gain

The connection between doping and gain is investigated using bias voltage scans of $1/C^2$ from C-V and the gain from charge collection on 50 μm thick LGAD sensors from the HPK prototype ECX20840 [9]. An explanation of the bias voltage steps with respect to the LGAD layer structure is given in the Appendix.

### 2.1 Capacitance-Voltage Scans

We investigated four sensor types, identical with the exception of the doping concentration which is varied in 10% steps. The capacitance C was measured at room temperature as a function of bias at 10 kHz and analysed in the form of $1/C^2$ vs. bias voltage scans (Fig. 1) to find the bias voltage to deplete the highly doped (p++) layer, $V_{GL}$, (also called the "Foot"). The importance of $V_{GL}$ is that it is linearly correlated with the relative doping concentration of the gain layer Fig. 1(right.)

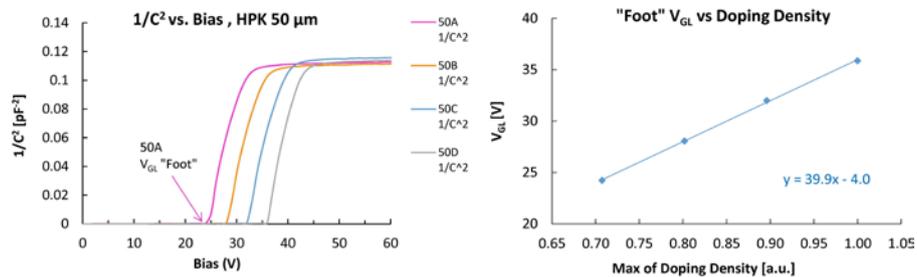

Fig.1 C-V analysis of the HPK prototype ECX20840 before radiation: (left): bias dependence of $1/C^2$ to find the bias voltage $V_{GL}$ which depletes the gain layer; (right): correlation between relative doping density and $V_{GL}$.

### 2.2 Charge Collection studies

Charge collection studies of the selected sensors have been reported before in Refs. [5, 6]. They use a telescope with β-particles from a $^{90}$Sr source triggered by a second fast LGAD. The collected charge of the device under test was measured with a custom high-bandwidth readout board as a function of bias. The gain was extracted by dividing the collected charge by the expected charge from a sensor without gain ("PiN") which pre-rad has a value close to 0.5 fC for 50 μm sensors, which is reduced during irradiation because of trapping [6, 10].

In Fig. 2 the results are shown for the gain vs. bias curves of the four sensors shown in the previous section; horizontal lines indicate the gain of G = 10 and 20, respectively. The corresponding bias voltages for a gain of G = 10 and 20, respectively, for the four sensors are shown in Fig 3 (left) as a function of the nominal doping density: a perfect linear correlation between V(G) and the doping concentration is observed, for both V(G=10) and V(G=20). The fact that, as shown in Fig. 3 (right), V(G=10) is linearly correlated with $V_{GL}$ as expected from the previous subsection is a confirmation of the internal consistency of the data.

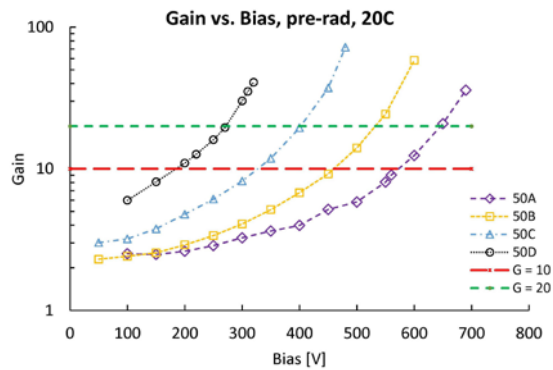

Fig.2 Charge collection for the HPK prototype ECX20840: gain vs. bias voltage

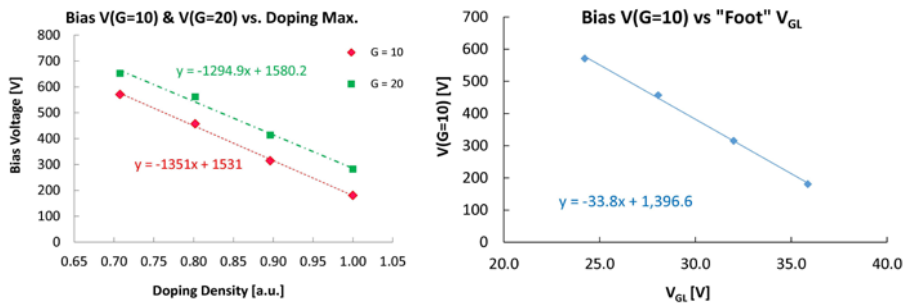

Fig. 3 Correlation between charge collection and doping for the unirradiated HPK prototypes ECX20840: (left): bias V(G=10) and V(G=20) vs. the relative doping density; (right): correlation between V(G=10) and $V_{GL}$.

The conclusion is that the gain-bias curves are correlated with the doping-bias curves. As a rule of thumb, the gain-bias curve shift by 13 V for a change of 1% in doping ( a general rule for LGADs) and by 34 V for a shift of 1 V in $V_{GL}$ ( which depends on the specific sensor). This hold also after irradiation, as demonstrated in Fig. 4 which shows the gain-bias curves for the lowest-doped sensor 50A before irradiation and for 50D before and after neutron irradiation to fluences of 3E14, 6E14, 1E15 Neq/cm$^2$. The pre-rad gain curve for 50A coincides with the curve for 50D after 6E14 Neq/cm$^2$ corresponding to a bias voltage shift of $\Delta V(G=10) = 420$ V, while from the difference in doping of 30% we expect a shift of $\Delta V(G=10) = 30*13 = 390$ V.

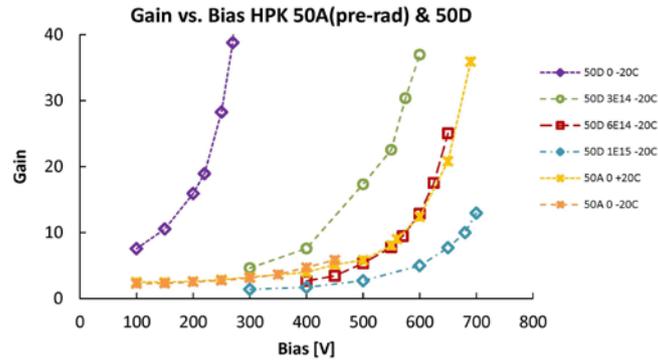

Fig.4 Gain-bias curves for HPK LGAD 50D and 50A pre-rad and 50D after neutron fluences of 3, 6, $10*10^{14}$ Neq/cm$^2$.

## 3. Acceptor Removal

In the following we consider the radiation dependence of the characteristics of three LGADs of close to ~ 50 µm thick FZ bulk and the same geometry of 1.3 x 1.3 mm$^2$ pad area serving as prototypes for the use in the High-Granularity Timing Detector (HGTD) of ATLAS [11] and the Endcap Timing Layer (ETL) of CMS [12], both upgrades for the HL-HLC [13]. Two HPK LGAD are from the combined ATLAS-CMS run EXX30330 and the FBK LGAD is from the INFN Torino run

UFSD3. They have different doping profiles as indicated by the $1/C^2$ vs. bias curves in Fig. 5 (left): the HPK-3.2 has the largest "Foot" and the deepest $p^{++}$ implant, the HPK-3.1 has a smaller "Foot" and shallower implant and the FBK UFSD3-C ( also called"FBK+C") has the smallest "Foot" and shallowest implant, yet is infused with Carbon, which has been shown to slow down the acceptor removal [7, 8, 14]. The doping density of the FZ bulk is of the order 3E12 $cm^{-3}$.

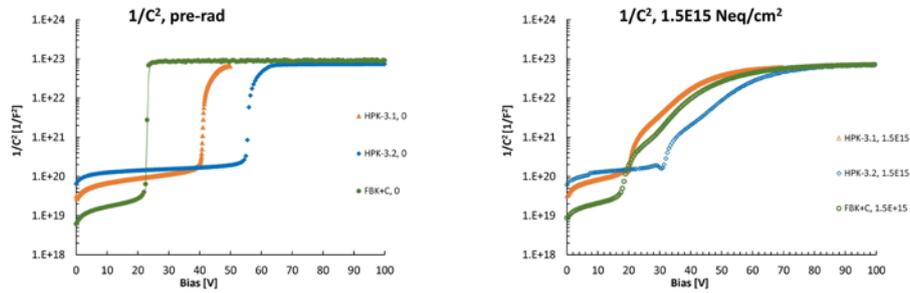

Fig. 5 Measured $1/C^2$ vs. bias curves for three selected sensors (left): before irradiation; (right): after a neutron fluence of 1.5E15 Neq/$cm^2$

After irradiation with neutrons at the TRIGA reactor in Ljubljana [15] the $1/C^2$ vs. bias curves show a shortening of the "Foot" and a decrease in the bulk resistivity (Fig. 5 (right), as indicated by the reduced slope of the curves in the bulk. As mentioned before, the gain of LGADs at fixed bias changes due to acceptor removal, which reduces $V_{GL}$. Its dependence on the fluence $\Phi$ is exponential with a coefficient c depending on the initial doping profile, the particle type and modifications to the bulk in the gain layer region:

$$V_{GL}(\Phi) = V_{GL}(0)*e^{(-c*\Phi)}$$

The measured fluence dependence of $V_{GL}$ for the three sensor types are shown in Fig 6 (left). The value of c differs between the sensor types:

c(HPK 3.1) = 1.6*c(HPK 3.2);

c(HPK 3.2) = 2.1*c(FBK+C)

These values are close to those reported in Ref [8].

Although before irradiation $V_{GL}$ is very different for the three sensors, in the fluence range of 3E15 Neq/$cm^2$ they become similar. Thus naively one would expect a similar bias dependence for their gain. In Fig. 6 (right) the bias dependence of the gain is shown at fluence of 3E15 Neq/cm$^2$: the values for $V_{GL}$ for (HPK 3.1), (FBK+C) and (HPK 3.2) are 10, 15, 20 V, respectively, while the bias voltages for a gain of 4, V(G=4), vary widely with 730, 500, 680 V, respectively, showing the advantage of the added carbon treatment for FBK+C. Clearly the naïve expectation is wrong.

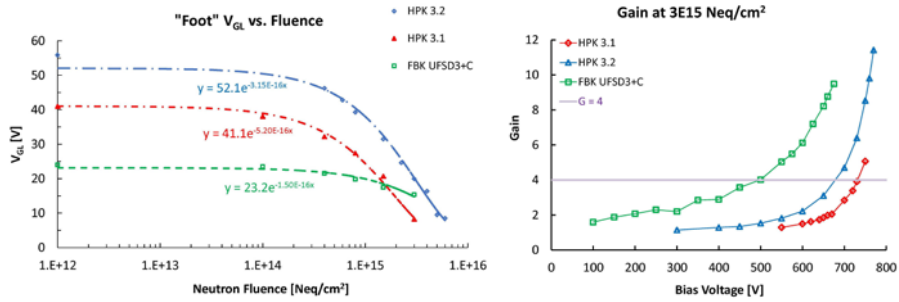

Fig. 6 (left): $V_{GL}$ as a function of neutron fluence for the three selected sensor types; (right): bias dependence of the gain for a neutron fluence of 3E15, with a gain G = 4 indicated.

## 4. Electric Field

The multiplication process is governed by the multiplication length $\lambda$. The number $\Delta N(\Delta x)$ of created carriers in a distance $\Delta x$, is

$$\Delta N(\Delta x) = e^{\Delta x/\lambda}$$

The multiplication length $\lambda$ shown in Fig. 9 as a function of the electric field depends approximately exponentially on the inverse of the electric field. For $\lambda = 1$ µm, the field needed for multiplication is about 270 kV/cm and the multiplication happens on the 1 µm scale.

The E-field in the multiplication process is supplied both by the gain layer and the bulk. The bulk, in addition to providing the active region for charge generation and the drift, contributes an E-field which after depletion depends linearly on the bias voltage. and thus can be investigated with charge collection studies measuring the bias to reach a fixed gain G, V(G). The gain layer adds an E-field proportional to the doping concentration which is investigated with C-V measurements measuring the voltage to deplete the gain layer, $V_{GL}$.

## 4.1 Electric field from gain layer and bulk

To show the two contributions to the field, a simple "Toy Simulation" of the E-field was done with two generic doping profiles of the gain layer on float zone (FZ) bulk. The one called "Deep" has a deep and narrow dopant distribution at about 2 μm depth and the other, called "Shallow", has shallower but wider profile at about 1.5 μm depth. One condition on the doping profiles of the doping layer is that the breakdown voltages of the two sensors match at about 200 V. Fig 7 shows the electric field simulated assuming a bias voltage close to the breakdown voltage before irradiation. The different contribution to the field from the bulk doping and from the gain layer (GL) doping are indicated, showing that pre-rad, the total field in the gain layer is 90% due to the high doping of the gain layer and only in a small part (10%) due to the highly resistive bulk of 50 μm depth. The deep implant offers the advantage of an extended high E-field close to the junction where the multiplication can occur.

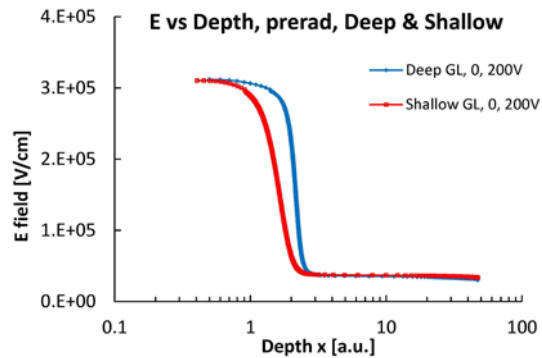

Fig. 7 "Toy" simulation of the pre-rad electric field vs. depth for two "typical" LGAD with deep and shallow gain layer doping, respectively, with the bias close to breakdown at 200 V.

## 4.2 Electric field after irradiation.

The simplified assumption concerning the effects of neutron irradiation are that the bulk doping increases linearly with fluence $\Phi$ by $0.02*\Phi$ [7], while the doping of the gain layer gets reduced by a factor $\exp(-c*\Phi)$, where c depends on the sensor, as shown in Section 3. The effect of radiation on the E-field is shown in Fig. 8 (left): a decreased gain layer contribution is only partially offset by an increased bulk contribution, but the pre-rad field can not be matched because the bias and thus the bulk field are limited by the breakdown of the sensor. For heavily irradiated sensors (Fig. 8 (right)) the bias voltage reach allows the field in the bulk to almost compensate for the loss of doping in the gain layer.

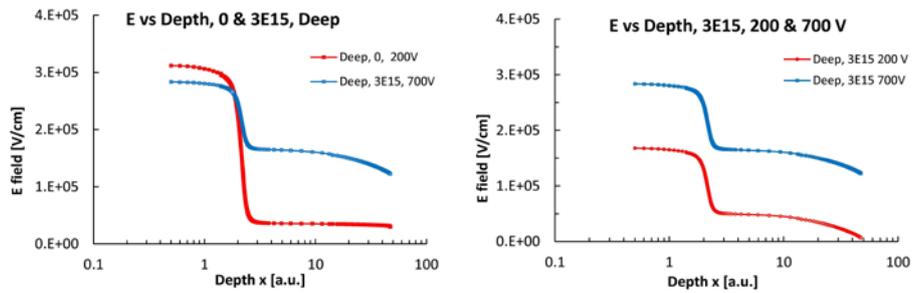

Fig. 8 Effect of irradiation on the electric field in the LGAD with deep gain layer: (left): pre-rad vs. high fluence with bias close to breakdown; (right): low bias vs. high bias for the "Deep" sensor heavily irradiated with neutrons to 3E15 Neq/ cm$^2$.

## 4.3 Multiplication Length $\lambda$ after irradiation

Since neutron radiation with fluence up to 3E15 Neq/cm$^2$ changes the balance of the electric field in the gain layer vs. in the bulk, we can look at the difference in gain in the two parts of the LGAD. Since the achievable maximum field is much higher in the gain layer (~300 kV/cm) than in the bulk (~170 kV/cm), the values of the multiplication length between the gain layer and the bulk differ by a factor 10 as shown in Fig. 9, a WF2 simulation [10] of the multiplication length $\lambda$ as a function of electric field following the model of Massey [16]. This assumes that impact ionization in the investigated fluence range is about the same as before irradiation, which is based on the observation that the mobility in that fluence range is not heavily affected. Therefore, the amount of radiation induced scattering centers is not very different than before irradiation.

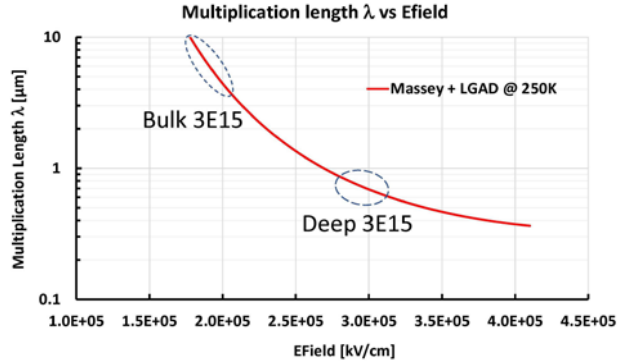

Fig. 9: WF2 simulation [10] of the multiplication length λ as a function of electric field following Ref [16]. Both the field achievable in the bulk and in the gain layer are shown for the "Deep" profile after irradiation with 3E15 Neq.

## 5. Correlation $V_{GL}$ – V(G=8)

As shown in the preceding sections, the "Foot" voltage $V_{GL}$ depends on the depth of the implant and doping of the gain layer, which establishes a highly localized electric field proportional to the amount of doping. Since these quantities are different for the different sensor types, the "Foot" voltage is sensor dependent. After irradiation the gain layer doping is reduced. Since the geometry of the implant is only slightly affected by irradiation, the "Foot" voltage, which is proportional to the electric field established by the gain layer, decreases. When a large voltage is applied to the sensor (much larger than the "Foot" voltage) the electric field in the gain layer, which is a superposition of the fields from the resistive bulk and the gain layer implant, increases. In this way the increase in detector voltage can compensate for a decrease in boron doping. A given signal gain corresponds to a given electric field in the gain layer, with components proportional to the "Foot" voltage and bulk detector voltage. This results in a linear relation between the "Foot" voltage and detector bias voltage for any given fixed signal gain.

This correlation is shown in Fig. 10 where for the three sensors considered the bias voltage for a gain of 8, V(G=8), is shown for the corresponding "Foot" voltage $V_{GL}$ for a variety of neutron fluences. The expected linear relationship between the two variables is observed, but also a difference in the dependence of the required bias voltage on the initial value of the "Foot" value (c.f. Section 3).

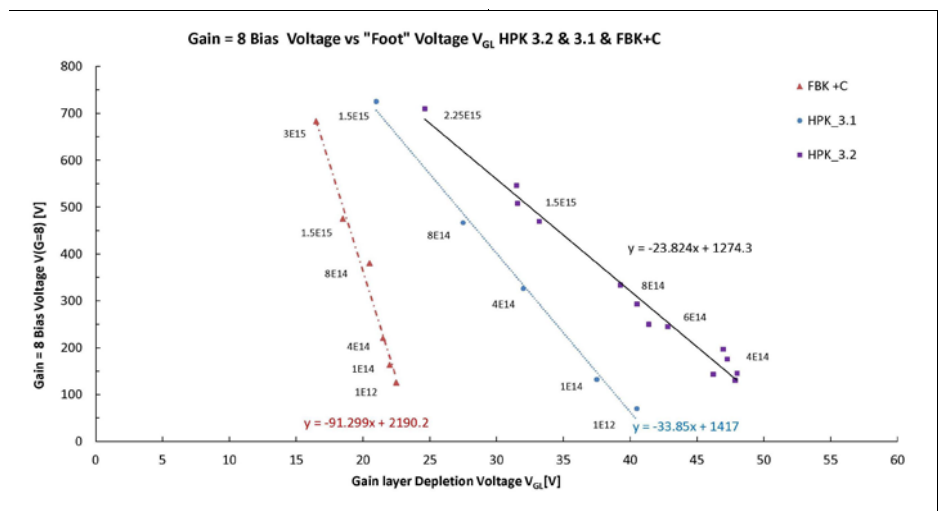

Fig. 10 Correlation between the bias for gain G=8, V(G=8), and the "Foot" voltage $V_{GL}$ for the three selected sensors.

Since at high fluences the small $V_{GL}$ can be measured with low-bias C-V measurements, yet V(G=8) can not be measured because of the breakdown of the sensors at 750V, one is tempted to make use of this relationship to extend it beyond the measured values of V(G=8) to predict the values of bias voltages required for higher fluence. With a caveat that this would require that the gain is still governed by the gain layer doping, which may not be, an extrapolation to higher fluences gives a value of V(G=8) in excess of 1100 V for HPK 3.1 at 3E15 Neq/cm$^2$, and in excess of 1000 V for HPK 3.2 at 6E15 Neq/cm$^2$, much larger than the breakdown voltage of 750 V for both sensors.

## 6. Conclusions

We have shown linear correlations before and after irradiation between several parameters governing the gain of Low-gain Avalanche Detectors (LGAD) being developed for large scale application in experimental particle physics.

The correlation between three parameters, a) the doping concentration of the gain layer, b) the bias voltage to deplete the gain layer, $V_{GL}$, and c) the bias voltage to reach a certain gain G, V(G), (or collected charge, CC, V(CC)), can then be used interchangeably to characterize the status of the gain, thus giving the experimenter the choice between different experimental procedures.

While V(G) requires a charge collection study at low temperature with high-voltage biasing and is limited by the breakdown voltage of the sensor, $V_{GL}$ can be measured at room temperature with low-bias C-V scans at 1kHz, which makes this method available for easy application.

We emphasize the advantage of the latter for large-scale quality monitoring of radiation campaigns.

## Acknowledgements

This work was supported by the United States Department of Energy, grant DE-FG02-04ER41286.

Part of this work has been financed by the European Union's Horizon 2020 Research and Innovation funding program, under Grant Agreement no. 654168 (AIDA-2020) and Grant Agreement no. 669529 (ERC UFSD669529), and by the Italian Ministero degli Affari Esteri and INFN Gruppo V.

This work was supported by the Slovenian Research Agency (project J1-1699 and program P1-0135)

This work was partially performed within the CERN RD50 collaboration [17].

## Appendix: LGAD structure and Capacitance

The LGAD used have a two-layered structure: a highly doped gain layer of about $x_{GL}$ = 1-2 μm thickness and a resistive bulk of about $x_B$ = 50 μm thickness. The doping in the gain layer $N_{GL}$ is of the order of $N_{GL} \approx 10^{16}$ cm$^{-3}$, which leads to a high electric field gradient across the layer, as shown in Section 4. The bulk has much lower doping density $N_B$, in our case $N_B \approx$ few $10^{12}$ cm$^{-3}$, and the electric field rises linearly from the backside to the gain layer for a uniform doping profile. Within the gain layer, the electric field needed for multiplication is the sum of the two contributions. Displacing radiation with hadrons impacts the doping in the two layers differently: the doping in the gain layer decreases due to acceptor removal, and the one in the bulk increases due to acceptor generation (see Fig. 20 of Ref. 6). Thus the ratio between the contributions to the gain from gain layer and bulk will decrease with increasing fluence.

The structure of LGADs is reflected in capacitance – bias voltage (C-V) scans. The depletion voltage for the two parts of the structure is given by

$$V_{GL} = N_{GL}/(2\varepsilon) \cdot x_{GL}^2, \quad V_B = N_B/(2\varepsilon) \cdot x_B^2 \quad , \qquad (A.1)$$

and the full depletion voltage of the LGAD $V_{FD}$ is thus $V_{FD} = V_{GL} + V_B$. A consequence of eq. A.1 is the relatively large gain layer depletion voltage $V_{GL}$ due to the large doping $N_{GL}$: before radiation, $V_{GL} > V_B$. In addition, with irradiation, $V_{GL}$ decreases and $V_B$ increases, as shown in Fig. 5. During the depletion of the gain layer, the capacitance is relatively high due to the small thickness, and only after the depletion of the gain layer does the capacitance start the well-known linear increase in the 1/C^2 vs bias plot observed in silicon sensors without gain. Thus $V_{GL}$ is called the "foot" and is defined by this transition as shown in Fig. 1. Even after irradiation, when the depletion of the bulk requires much larger bias voltages, $V_{GL}$ can be determined well as shown in Fig. 5.

The above equations assume constant doping density, but the result is more general, that is the foot idea works even without uniform doping and even with gaps in the doping.